\documentclass{emulateapj}
\usepackage{amssymb}
\usepackage{epsfig}
\usepackage{amsmath}
\usepackage{enumerate}
\usepackage{natbib}
\usepackage[breaklinks,colorlinks,citecolor=blue]{hyperref}
\usepackage{ulem}
\usepackage{color}

\shorttitle{ }
\shortauthors{ }

\begin{document}

\title{A Pulsar Wind Nebula Model Applied to Short GRB 050724}
\author{Wei-Li Lin\altaffilmark{1,2}, Ling-Jun Wang\altaffilmark{3}, Zi-Gao
Dai\altaffilmark{1,2}}

\begin{abstract}
A subset of short gamma-ray bursts (sGRBs) have been found to be
characterized by near-infrared/optical bumps at $\sim 1$ days, some of which
exhibit almost concurrent X-ray flares. Although the near-infrared/optical
bumps may be a signature of kilonovae, the
X-ray flares are not consistent with kilonovae. It is widely believed that
sGRBs are produced by the mergers of double compact objects, during which
sub-relativistic ejecta are launched. In this paper we propose that the
above optical/X-ray features are indicative of the formation of long-lived
magnetars following the mergers of double neutron stars. Observations and
theoretical works imply that the spin-down power of the magnetars is
injected into the ejecta as ultra-relativistic electron-positron pairs, i.e.
pulsar wind nebulae (PWNe). Here we suggest such a PWN model and find that
the optical bump and X-ray flare observed in GRB 050724 can be well
understood in this PWN model. We show that the optical bump and X-ray flare
may have different origins. Our results strengthen the evidence for the
formation of magnetars in double neutron star mergers and justify the
validity of the PWN model.
\end{abstract}


\affil{\altaffilmark{1}School of Astronomy and Space Science, Nanjing
University, Nanjing 210093, China; dzg@nju.edu.cn}

\affil{\altaffilmark{2}Key Laboratory of
Modern Astronomy and Astrophysics (Nanjing University),
Ministry of Education, Nanjing 210093, China}

\affil{\altaffilmark{3}Astroparticle Physics, Institute of High Energy Physics,
Chinese Academy of Sciences, Beijing 100049, China}


\section{Introduction}

\label{sec: introduction}

Short gamma-ray bursts (sGRBs) have been widely believed to originate from
the mergers of neutron star - neutron star (NS-NS) or neutron star - black
hole (NS-BH) systems \citep{Eichler1989, Narayan1992, Berger14}. A
recent gravitational-wave (GW) event (GW170817), has been
identified as the signal from a binary neutron star inspiral %
\citep{Abbott2017PhRvL.119p1101A}. The association of GW170817 with an
sGRB (GRB 170817A)
\citep{Goldstein2017ApJ...848L..14G,
Savchenko2017ApJ...848L..15S}, provides a direct evidence for the binary NS
coalescences as progenitors of at least part of sGRBs.\footnote{%
Independent of a kilonova that can interpret the rapid evolution in
ultraviolet/optical/near-infrared bands %
\citep[e.g.,][]{Evans2017arXiv171005437E, Villar2017ApJ...851L..21V}, both
the X-ray and radio observations of GRB 170817A are consistent with either a
mildly relativistic cocoon or an off-axis relativistic jet
\citep[e.g.,][]{Hallinan2017arXiv171005435H, Margutti2017ApJ...848L..20M,
Troja2017Natur.551...71T, Xiao2017ApJ...850L..41X}.}

The products of the NS-NS or NS-BH mergers are usually BHs. The outcome of
NS-NS mergers could also be highly magnetized NSs
\citep[magnetars;][]{Dai+Lu1998A&A...333L..87D, Dai98b, Zhang+Meszaros2001ApJ...552L..35Z,
Giacomazzo13, Giacomazzo15}. Direct identification of the merger product is
at the reach of current advanced GW detectors. Although the mergers of NS-NS
or NS-BH can be discerned by their different GW signals \citep{Bartos13},
the combined detection of GW signals and their electromagnetic (EM)
counterparts\ can improve our understanding of the physics of the merger
process \citep{Fernandez16}, which can be a probe of the long-pursuing
equation of state for supra-nuclear matter \citep{Paschalidis17, Radice17}.

The most energetic EM counterparts are sGRB prompt emission. If the merger
product is a differentially rotating-supported supramassive magnetar that
eventually collapses into a BH, the sGRB afterglow is characterized by
shallow decay followed by steep decline \citep{Rowlinson10, Rowlinson13}.
However, sGRB and its immediate afterglow may not be the feasible candidate
for the combined detection of a GW signal because of its low rate within the
GW detection horizon due to the relativistic beaming effect.

Fortunately, there are a variety of almost isotropic EM counterparts that
are bright enough to be easily detected. The NS magnetic interaction %
\citep{Lai2012ApJ...757L...3L, Piro12} with the companion NS or BH produces
precursor emission in radio \citep{Wang2016ApJ...822L...7W} and X-ray %
\citep{Palenzuela13} bands. Sub-relativistic outflows would be launched
by the mergers of NS-NS or NS-BH systems. The interaction of such outflows
with surrounding interstellar medium (ISM) produces radio flares that are
detectable for weeks \citep{Nakar11}.

If the merger product is a BH, the remnant could emerge as a
kilonova heated by the radioactive $r$-process material
\citep{Li+Paczynski1998ApJ...507L..59L, Metzger2010MNRAS.406.2650M, Dietrich-etal17,
Fernandez17, Metzger17}. The evidence for such kilonovae boosted recently by
the detection of near-infrared/optical bumps at $\sim 1-10$ days following
some sGRBs
\citep{Berger2013ApJ...774L..23B, Tanvir2013Natur.500..547T,
 Yang2015NatCo...6E7323Y, Jin2016NatCo...712898J}.\footnote{%
The early X-ray light curve of GRB 130603B was attributed to a short-lived
magnetar by \cite{Fan2013ApJ...779L..25F}.}

If, on the other hand, the merger product is a magnetar, there could be very
rich EM counterparts. The magnetic dissipation by a differentially rotating
millisecond pulsar born after NS-NS merger could produce early-time multiple
X-ray flares \citep{Dai2006Sci...311.1127D}. The continuous energy injection
by a stable post-merger magnetar will boost the luminosity of the
radioactively powered kilonova to become a luminous mergernova %
\citep{Yu2013ApJ...776L..40Y} that lasts for $\sim 10$ days in
optical/ultraviolet bands. The interaction (forward shock) of the
magnetar-aided ejecta with ISM will produce multi-band EM emission %
\citep{Gao2013ApJ...771...86G, Yu2013ApJ...776L..40Y}.

Some sGRBs (e.g., GRB 050724, GRB 061006, GRB 070714B and GRB 080503) show
optical bumps a few days in length, a few of which (GRB 050724 and GRB 080503)
were accompanied by almost concurrent X-ray flares. Different models were
put forward to account for these features. For example, \cite%
{Panaitescu2007MNRAS.379..331P} proposed that the X-ray flare in the
afterglow light curve of GRB 050724 was caused by an energy injection.
\citet{Gao2015ApJ...807..163G,
Gao2017ApJ...837...50G} attributed the late optical re-brightenings to the
mergernova emission. To account for their concurrent X-ray flares,
\citet{Gao2015ApJ...807..163G,
Gao2017ApJ...837...50G} proposed that the X-ray emission is a leakage of the
spin-down luminosity of the magnetars.

An alternative scenario
\citep{Wang+Dai2013, Wang+Dai2015,
Wang2016ApJ...823...15W} about the multi-band emission of the
magnetar-powered post-merger systems is based on the observations %
\citep{Gaensler06, Hester08} and theories %
\citep{Rees74,Kennel84a,Kennel84b,Begelman92, Chen17} of pulsar wind nebulae
(PWNe), which are generated by the complex interactions between the magnetar
wind, ejecta, and ambient ISM. The model, different from the above
mergernova model \citep{Yu2013ApJ...776L..40Y, Gao2015ApJ...807..163G}, is
built on the basis that the magnetar wind is Poynting flux dominated
initially and then converted into energy flux of ultra-relativistic
electron-positron pairs ($e^{+}e^{-}$ pairs;
\citealt{Coroniti1990,
Michel1994, Dai2004, Yu+Dai2007A&A...470..119Y}, see also %
\citealp{Geng2016ApJ...825..107G} for the applications to GRB jets).%
\footnote{%
The effect of wind magnetization is studied by \citet{Liu2016A&A...592A..92L}
based on the work of \citet{Zhang+Kobayashi2005ApJ...628..315Z}.}

\begin{figure}[tbp]
\centering\resizebox{1\hsize}{!}{
\includegraphics{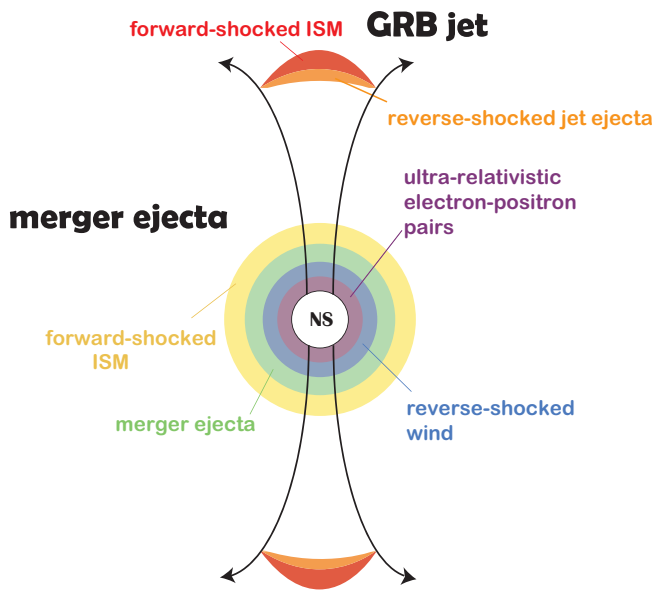}}
\caption{Schematic diagram of the evolution of an sGRB and a PWN after the
formation of a post-merger millisecond magnetar.}
\label{fig:sketch}
\end{figure}

Here we follow the PWN model \citep{Wang+Dai2015}. In this model, the ejecta
accelerated by the pressure of the magnetar wind sweeps up the ambient
medium to form a forward shock. The interaction between the ejecta and
magnetar wind forms a lepton-dominated reverse shock, as schematized in
Figure \ref{fig:sketch}. In the original model \citep{Wang+Dai2015} the
energy loss due to the reverse-shock emission was ignored, as we always do
in the modeling of GRB afterglow. However, because the reverse shock is
lepton-dominated, the shock energy is carried mainly by leptons, whose
energy loss will have a significant impact on the shock dynamics.

In this work, the energy loss due to the reverse-shock emission is
considered. With such a revised PWN model, together with the GRB afterglow
model, we interpret the X-ray flare at $\sim 1$ day after the trigger of GRB
050724 as originating from the PWN reverse-shock emission, and the optical
bump originating from the ejecta thermal emission. This paper is organized
as follows. In Section~\ref{sec: model description}, we introduce the GRB
afterglow and PWN models used in this work. In Section~\ref{sec: application}%
, we show the fitting results to multi-band observations of GRB 050724.
Finally, we discuss and conclude our findings in Section~\ref{sec: summary}.

\section{Model description}

\label{sec: model description}

\subsection{GRB afterglow}

\label{subsec: GRB afterglow}

The merger of NS-NS binary triggers the launch of a collimated relativistic
jet. This jet gives rise to a forward shock that moves outward into the ISM
and a reverse shock that backs into the unshocked ejecta (e.g.,
\citealp{Sari+Piran1995, Sari+Piran1999, Kobayashi2000,
Kobayashi+Sari2000}). Listed here are four regions of interest, namely, the
unshocked ambient medium (region 1), forward-shocked medium (region 2),
reverse-shocked ejecta (region 3) and unshocked ejecta (region 4). In a
similar way as used by \citet{Huang2000} and \citet{Lan2016}, the non-rest
energy of this system can be expressed as
\begin{equation}
\begin{split}
E_{\mathrm{j}}=& (\gamma _{\mathrm{j}}^{2}-1)M_{\mathrm{sw},\mathrm{j}%
}c^{2}+(\gamma -1)M_{3,\mathrm{j}}c^{2}+\gamma _{\mathrm{j}}U_{3,\mathrm{j}%
}^{\prime }+ \\
& (\gamma _{4,\mathrm{j}}-1)(M_{\mathrm{ej},\mathrm{j}}-M_{3,\mathrm{j}%
})c^{2},
\end{split}
\label{eq: E_tot_ag}
\end{equation}%
where $\gamma _{\mathrm{j}}$ is the bulk Lorentz factor of the jet, $\gamma
_{4,\mathrm{j}}$ is the Lorentz factor of region 4, $M_{\mathrm{sw},\mathrm{j%
}}$ is the mass of swept-up ambient medium (with number density $n$), $M_{%
\mathrm{ej},\mathrm{j}}$ is the mass of ejecta (i.e., initial mass of the
jet), $M_{3,\mathrm{j}}$ is the mass of reverse-shocked ejecta, and $U_{3,%
\mathrm{j}}^{\prime }$ is the internal energy of region 3 measured in the
co-moving frame.

Energy conservation yields
\begin{equation}
\dfrac{\mathrm{d}\gamma _{\mathrm{j}}}{\mathrm{d}t}=-\dfrac{\gamma _{\mathrm{%
j}}\mathcal{D}\dfrac{\mathrm{d}U_{3,\mathrm{j}}^{\prime }}{\mathrm{d}%
t^{\prime }}/c^{2}+(\gamma _{\mathrm{j}}^{2}-1)\dfrac{\mathrm{d}M_{\mathrm{sw%
},\mathrm{j}}}{\mathrm{d}t}+(\gamma _{\mathrm{j}}-\gamma _{4,\mathrm{j}})%
\dfrac{\mathrm{d}M_{3,\mathrm{j}}}{\mathrm{d}t}}{U_{3,\mathrm{j}}^{\prime
}/c^{2}+2\gamma _{\mathrm{j}}M_{\mathrm{sw},\mathrm{j}}+M_{3,\mathrm{j}}}.
\label{eq: dynamics_ag}
\end{equation}%
where the co-moving-frame time $t^{\prime }$ is equal to observer's time $t$
multiplied by the Doppler factor $\mathcal{D}$.

Before the reverse shock crosses region 4, the evolution of $U_{3,\mathrm{j}%
}^{\prime }$ depends mostly on the reverse shock process, (i.e., $\mathrm{d}%
U_{3,\mathrm{j}}^{\prime }=(\gamma _{34,\mathrm{j}}-1)\mathrm{d}M_{3,\mathrm{%
j}}c^{2}$, where $\gamma _{34,\mathrm{j}}$ is the Lorentz factor of the jet
measured in the frame of region 4). However, after the cross time $t_{%
\mathrm{c}}$, region 3 evolves in an adiabatic expansion, where we adopt
(see also Equation~$\left( 7\right) $ of \citealp{Lan2016})
\begin{equation}
\dfrac{\mathrm{d}U_{3,\mathrm{j}}^{\prime }}{\mathrm{d}t^{\prime }}=-\dfrac{1%
}{3}\dfrac{e_{3,\mathrm{j}}^{\prime }(t_{\mathrm{c}})V_{3,\mathrm{j}%
}^{\prime \frac{4}{3}}(t_{\mathrm{c}})}{V_{3,\mathrm{j}}^{\prime \frac{4}{3}}%
}\dfrac{\mathrm{d}V_{3,\mathrm{j}}^{\prime }}{\mathrm{d}t^{\prime }},
\label{eq: dU3_1_ag}
\end{equation}%
with $V_{3,\mathrm{j}}^{\prime }$ and $e_{3,\mathrm{j}}^{\prime }$ being the
co-moving-frame volume and internal energy density of region 3, respectively.

The calculation of other relevant quantities is presented in Appendix~\ref%
{app: calculation of jets}. In the context of GRB afterglow, we do not
consider the effect of radiative loss on dynamics because the energy
fraction of the shocked electrons is relatively low.

\subsection{PWN powered by the magnetar wind}

\label{subsec: PWN}

The coalescence of binary NSs ejects sub-relativistic and near-isotropic
outflows with masses $10^{-4}-10^{-2}M_{\odot }$
\citep{Rezzolla10, Bauswein13,
Hotokezaka2013PhRvD..87b4001H, Baiotti17, Ciolfi17}. If the merger product
is a stable magnetar, the Poynting flux in the immediate vicinity of the
magnetar is converted to $e^{+}e^{-}$ pairs that power the ejecta to form a
PWN \citep{Gaensler06, Aharonian12}. The merger ejecta are sandwiched
between the ambient medium and magnetar wind and become a thin layer. The
evolution of the PWN is shown in Figure \ref{fig:sketch}. In this diagram,
there are five regions, including the unshocked ambient medium (region 1),
shocked medium (region 2), merger ejecta (ejecta region), shocked magnetar
wind (region 3) and unshocked magnetar wind (region 4).

We investigate the dynamical evolution of the ejecta powered by a newborn
millisecond magnetar wind (i.e., ultra-relativistic $e^{+}e^{-}$ pairs). The
spin-down luminosity carried by the wind is $L_{\mathrm{sd}}=L_{0}(1+t/t_{%
\mathrm{sd}})^{-2}$ with $L_{0}\approx
10^{47}B_{14}^{2}R_{6}^{6}P_{0,-3}^{-4}~\mathrm{erg\ s^{-1}}$ and spin-down
timescale $t_{\mathrm{sd}}\approx 2\times
10^{5}I_{45}B_{14}^{-2}P_{0,-3}^{2}R_{6}^{-6}$ s, where the magnetar is
characterized by the following parameters: the surface magnetic field
strength $B$, initial spin period $P_{0}$, moment of inertia $I$ and stellar
radius $R$. Here we adopt the convention $Q=10^{n}Q_{n}$ in cgs units and
set $I_{45}=1.5$ and $R_{6}=1$.

The total energy of this system with Lorentz factor $\gamma $ can be written
as
\begin{equation}
E=\gamma ^{2}M_{\mathrm{sw}}c^{2}+\gamma M_{\mathrm{ej}}c^{2}+\gamma
M_{3}c^{2}+\gamma (U_{\mathrm{ej}}^{\prime }+U_{3}^{\prime }).
\label{eq: E_tot}
\end{equation}%
Analogous to the notation in Section~\ref{subsec: GRB afterglow}, $\gamma
_{4}$ is the Lorentz factor of the region 4, $M_{\mathrm{sw}}$ is the mass
of forward-shocked ambient medium, $M_{\mathrm{ej}}$ is the mass of merger
ejecta, $M_{3}$ is the mass of reverse-shocked magnetar wind, and $U_{%
\mathrm{ej}}^{\prime }$ and $U_{3}^{\prime }$ are the internal energy of
corresponding regions in the co-moving frame. As seen in this expression,
instead of considering the total energy evolution of region 3 %
\citep{Wang+Dai2015}, we list separately kinetic and internal energy of
region 3.

In light of the energy conservation, we can obtain the dynamics as
\begin{equation}
\dfrac{\mathrm{d}\gamma }{\mathrm{d}t}=\dfrac{\left[ \dfrac{\mathrm{d}E}{%
\mathrm{d}t}-\gamma \mathcal{D}\left( \dfrac{\mathrm{d}U_{\mathrm{ej}%
}^{\prime }}{\mathrm{d}t^{\prime }}+\dfrac{\mathrm{d}U_{3}^{\prime }}{%
\mathrm{d}t^{\prime }}\right) \right] /c^{2}-\gamma ^{2}\dfrac{\mathrm{d}M_{%
\mathrm{sw}}}{\mathrm{d}t}-\gamma \dfrac{\mathrm{d}M_{3}}{\mathrm{d}t}}{%
\left( U_{\mathrm{ej}}^{\prime }+U_{3}^{\prime }\right) /c^{2}+2\gamma M_{%
\mathrm{sw}}+M_{\mathrm{ej}}+M_{3}}.  \label{eq: dynamics}
\end{equation}%
The internal energy of region 3 evolves as
\begin{equation}
\dfrac{\mathrm{d}U_{3}^{\prime }}{\mathrm{d}t^{\prime }}=\dfrac{\mathrm{d}U_{%
\mathrm{3,rs}}^{\prime }}{\mathrm{d}t^{\prime }}-L_{3}^{\prime
}-p_{3}^{\prime }\dfrac{\mathrm{d}V_{\mathrm{3,enc}}^{\prime }}{\mathrm{d}%
t^{\prime }}.  \label{eq: dU3_1}
\end{equation}%
The internal energy increment $\mathrm{d}U_{\mathrm{3,rs}}^{\prime }=\left(
\gamma _{34}-1\right) \mathrm{d}M_{3}c^{2}$ corresponds to the
instantaneously reverse-shocked $e^{+}e^{-}$. We demonstrate $\mathrm{d}U_{%
\mathrm{3,rs}}^{\prime }/\mathrm{d}t^{\prime }\sim L_{\mathrm{sd}}^{\prime }$
($=\mathcal{D}^{2}L_{\mathrm{sd}}$) for low bulk Lorentz factor in the
Appendix~\ref{app: dUrs}.

The radiation luminosity $L_{3}^{\prime }$ due to reverse-shock emission can
be expressed as
\begin{equation}
L_{3}^{\prime }=\epsilon _{\mathrm{r,rs}}\dfrac{\mathrm{d}U_{\mathrm{3,rs}%
}^{\prime }}{\mathrm{d}t^{\prime }},  \label{eq: L3_1}
\end{equation}%
where the radiative efficiency can be estimated by a function of the
electron equipartition factor $\epsilon _{\mathrm{e,rs}}$, and the
co-moving-frame timescales of expansion and synchrotron cooling ($t_{\mathrm{%
ex}}^{\prime }$ and $t_{\mathrm{syn}}^{\prime }$;
\citealt{Dai1999,
Huang2000}), i.e.,
\begin{equation}
\epsilon _{\mathrm{r,rs}}=\epsilon _{\mathrm{e,rs}}\dfrac{t_{\mathrm{syn}%
}^{\prime -1}}{t_{\mathrm{syn}}^{\prime -1}+t_{\mathrm{ex}}^{\prime -1}}.
\label{eq: epsilon_rs}
\end{equation}%
Because the reverse-shocked $e^{+}e^{-}$ pairs are in fast cooling regime,
radiative efficiency is determined by $\epsilon _{\mathrm{e,rs}}$. The high
value, $\epsilon _{\mathrm{e,rs}}=0.9$, as adopted by \citet{Dai2004},
indicates the significant radiation loss of region 3, which is ignored in %
\citet{Wang+Dai2015}. Consequently, reverse-shock emission comes from
leptons that just get heated by reverse shock, rather than all shocked
leptons in region 3.

As for the third term in Equation~(\ref{eq: dU3_1}), the
average pressure in region 3 is $p_{3}^{\prime }=U_{3}^{\prime
}/(3V_{3}^{\prime })$ and the co-moving volume enclosed by region 3,
different from the comoving volume of region 3 ($V_{3}^{\prime }$),
increases at
\begin{equation}
\dfrac{\mathrm{d}V_{\mathrm{3,enc}}^{\prime }}{\mathrm{d}t}=4\pi r^{2}\beta
c.  \label{eq: dV_3_enc}
\end{equation}

The evolution of the internal energy of ejecta is related to the absorbed
fraction of radiation luminosity from region 3, (i.e., $(1-e^{-\tau
})L_{3}^{\prime }$), the thermal energy loss is similar to %
\citet{Kasen+Bildsten2010}
\begin{equation}
L_{\mathrm{th}}^{\prime }=\dfrac{U_{\mathrm{ej}}^{\prime }c}{\max (1,\tau
)\Delta _{\mathrm{ej}}^{\prime }},  \label{eq: L_ej_1}
\end{equation}%
and the radioactive power \citep{Korobkin2012}
\begin{equation}
L_{\mathrm{ra}}^{\prime }=4\times 10^{49}M_{\mathrm{ej,-2}}\left( \dfrac{1}{2%
}-\dfrac{1}{\pi }\arctan \dfrac{t^{\prime }-t_{0}^{\prime }}{t_{\sigma
}^{\prime }}\right) ^{1.3}\ \mathrm{erg\ s^{-1}},  \label{eq: L_ra_1}
\end{equation}%
where $t_{0}^{\prime }=1.3$ s, $t_{\sigma }^{\prime }=0.11$ s and $\tau
=\kappa (M_{\mathrm{ej}}/V_{\mathrm{ej}}^{\prime })\Delta _{\mathrm{ej}%
}^{\prime }$ with $\kappa $ being the opacity. In other words, we have
\begin{equation}
\dfrac{\mathrm{d}U_{\mathrm{ej}}^{\prime }}{\mathrm{d}t^{\prime }}%
=(1-e^{-\tau })L_{3}^{\prime }+L_{\mathrm{ra}}^{\prime }-L_{\mathrm{th}%
}^{\prime }.  \label{eq: dUej}
\end{equation}%
The work that the ejecta does on region 2 is counteracted by that region 3
material does on ejecta, so the total effect of these two works is neglected
here (see detailed discussion in \citealt{Wang+Dai2015}).

A combination of energy evolution in every region gives the energy variation
of the whole system, i.e.,
\begin{equation}
\dfrac{\mathrm{d} E}{\mathrm{d} t} = L_{\mathrm{sd}} + \mathcal{D}^2
\left(L_{\mathrm{ra}}^{\prime}- L_{\mathrm{th}}^{\prime}- L_{3}^{\prime
}e^{-\tau}\right)+ \dfrac{\mathrm{d} M_{\mathrm{sw}}}{\mathrm{d} t}c^2.
\label{eq: dE_tot}
\end{equation}

\begin{figure}[tbp]
\centering\resizebox{1\hsize}{!}{
\includegraphics{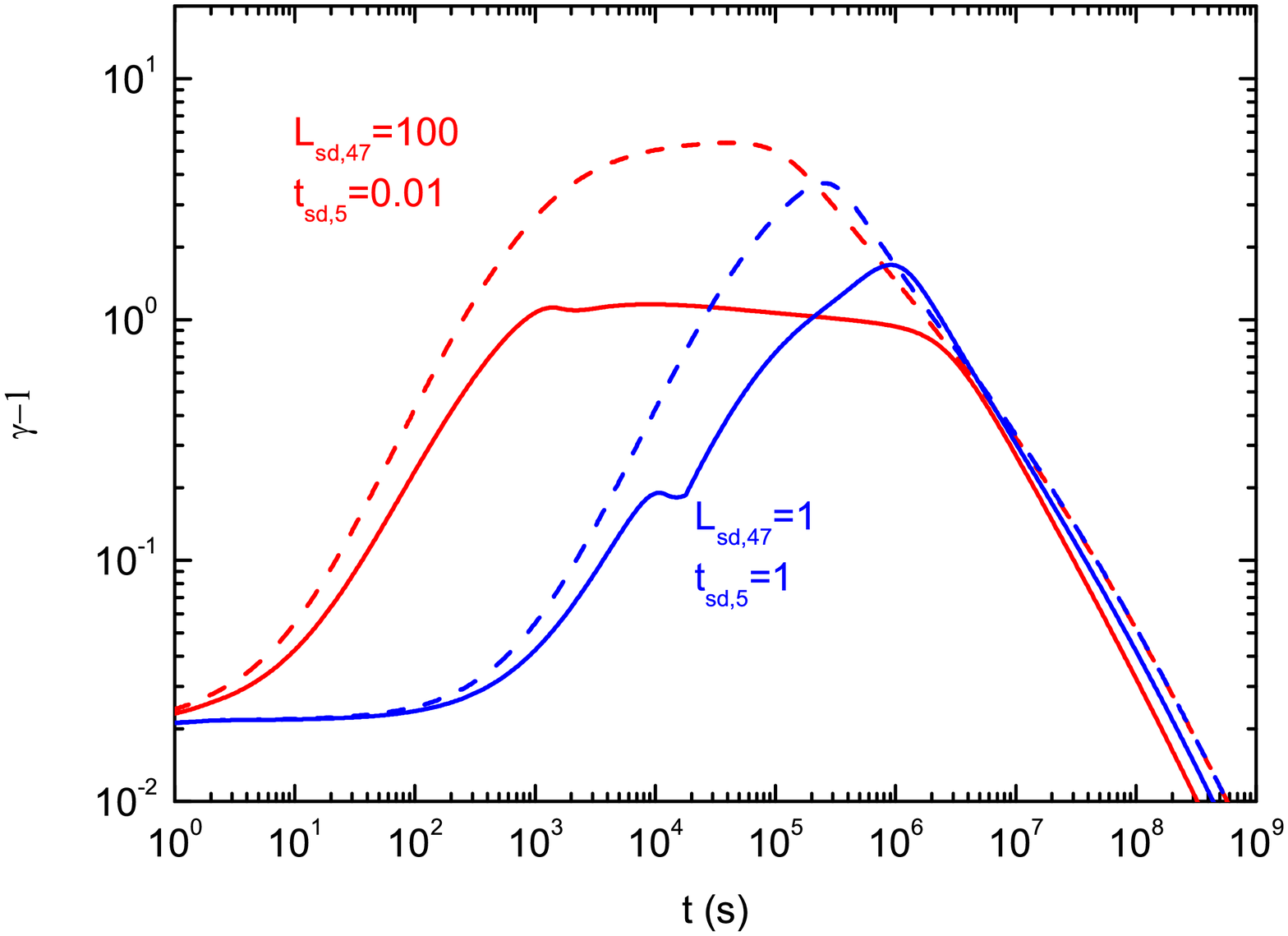}}
\caption{Dynamics derived from our PWN model (solid line) compared with
results without consideration of radiation loss of region 3 (i.e., setting $%
\protect\epsilon _{\mathrm{r,rs}}=0$; dashed line). We also show the results
of $L_{\mathrm{sd}}=10^{47}~\mathrm{erg\ s^{-1}}$ with $t_{\mathrm{sd}%
}=10^{5}$ s (blue line) and $L_{\mathrm{sd}}=10^{49}~\mathrm{erg\ s^{-1}}$
with $t_{\mathrm{sd}}=10^{3}$ s (red line), given that initial speed $%
v_{i}=0.2c$, $M_{\mathrm{ej}}=10^{-3}~M_{\odot }$, $n_{1}=0.1~\mathrm{cm^{-3}}$,
$\protect\gamma %
_{4}=10^{4}$ and $\protect\kappa =1~\mathrm{cm^{2}\ g^{-1}}$.}
\label{fig: dynamics-mg-comparions}
\end{figure}

Based on the above differential equation, we can obtain the dynamics of
merger ejecta system. Figure~\ref{fig: dynamics-mg-comparions} displays the
dynamical evolution of our model with and without consideration of radiation
loss of region 3. In this model, a reverse shock is generated when
the ultra-relativistic pulsar wind catches up with the sub-relativistic
ejecta. Most of the spin-down energy is at first transferred into the
internal energy of reverse-shocked wind. Thus, at early times, its pressure
is much higher than that of forward-shocked ISM and contributes to the
acceleration of the whole system. However, because of radiation loss, the
magnetar wind comprised of $e^{+}e^{-}$ pairs cannot accelerate the total
system efficiently. Moreover, due to the absorption and weak thermal
emission of ejecta in the early time, a large amount of radiation energy is
deposited into the internal energy of ejecta, which demands large energy
budget for acceleration and in a way, slows down the speed-up process.

In addition, the ejecta system could be accelerated continuously
even after $t_{\mathrm{sd}}$ (see the blue solid line in Figure~\ref{fig:
dynamics-mg-comparions} as an example) because several factors may
contribute to the acceleration. The large amount of internal energy
deposited in region 3 and ejecta could continuously accelerate the system.
The radioactive decay of the material in ejecta could keep heating the
ejecta. These factors may dominate over the decelerating effects such as%
 the energy loss due to thermal emission of the ejecta for a while
after $t_{\mathrm{sd}}$. Figure~\ref{fig: dynamics-mg-comparions} also
indicates that for magnetars with the same rotational energy, the dynamics
show strong dependence on the spin-down luminosity and spin-down timescale,
which are determined by the spin period and magnetic field strength of the
central magnetar. Thus, it can be concluded that the characteristics of
central magnetar play an important role in the dynamics and consequently the
emission of PWN components.

The shocked regions (regions 2 and 3) are characterized by synchrotron
radiation, which is computed in a standard way \citep[e.g.,][]{Wang+Dai2015}%
. Compared with such synchrotron radiation, we consider thermal radiation in
the ejecta, of which the flux at an observed frequency $\nu $ can be
expressed as
\begin{equation}
F_{\nu ,\mathrm{th}}=\dfrac{1}{\max (1,\tau _{\nu })D_{\mathrm{L}}^{2}}%
\dfrac{2\pi \mathcal{D}^{2}r^{2}}{h^{3}c^{2}\nu }\dfrac{\Delta _{\mathrm{ej}%
}^{\prime }}{r}\dfrac{(h\nu /\mathcal{D})^{4}}{\exp (h\nu /\mathcal{D}%
kT^{\prime })-1},  \label{eq:F_nu-th}
\end{equation}%
where $\tau _{\nu }$ is the wavelength-dependent optical depth, $D_{\mathrm{L%
}}$ is the luminosity distance and $T^{\prime }=(e_{\mathrm{ej}}^{\prime
}/a)^{1/4}$ denotes the blackbody temperature. It should be noted that the
pressure of ejecta $p_{\mathrm{ej}}^{\prime }=e_{\mathrm{ej}}^{\prime }/3$
can be expressed as $p_{\mathrm{ej}}^{\prime }=p_{3}^{\prime }+(1-e^{-\tau
})L_{3}^{\prime }/(4\pi r^{2}c)$, where the second term is the radiation
pressure corresponding to the radiation luminosity $(1-e^{-\tau
})L_{3}^{\prime }$. Calculation of observed flux from both region 3 and
ejecta should take into account the absorption and scattering effects of
ejecta, so we denote the opacity $\kappa $ for optical band and $\kappa _{%
\mathrm{X}}$ for X-ray band.

\begin{figure}[tbp]
\centering\resizebox{1\hsize}{!}{
\includegraphics{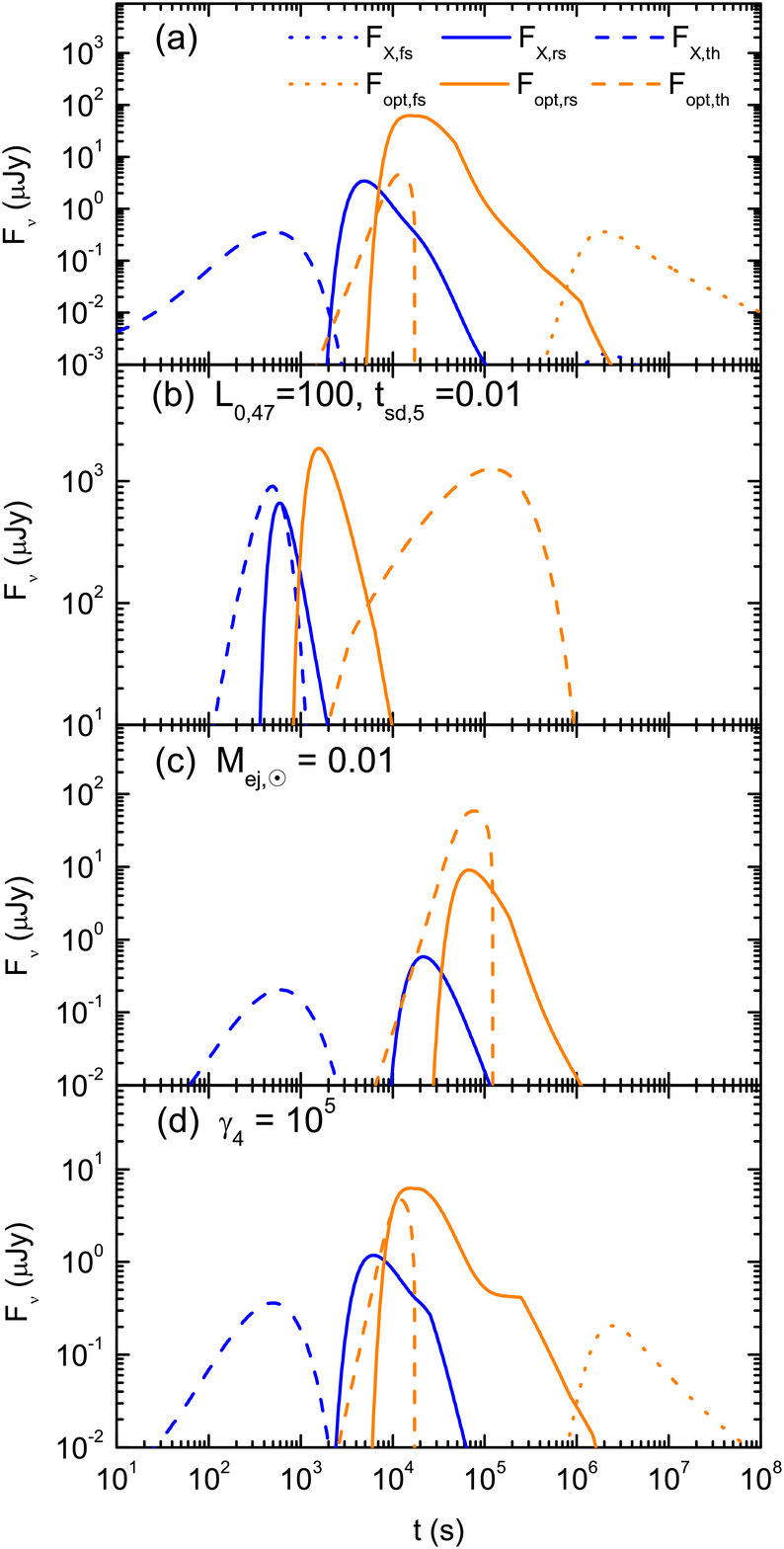}} 
\caption{X-ray and optical light curves sensitive to the parameters in our
PWN model. The dashed, dotted and solid lines correspond to the emission of
thermal, forward shock and reverse shock in PWN system, respectively. Panel (a)
results with parameters of redshift $z=0.2$, $n_{1}=0.1~\mathrm{cm^{-3}}$%
, $L_{\mathrm{sd}}=10^{47}~\mathrm{erg\ s^{-1}}$, $t_{\mathrm{sd}}=10^{5}$
s, $M_{\mathrm{ej}}=10^{-3}~M_{\odot }$, $v_{i}=0.2c$, $\protect\gamma %
_{4}=10^{4}$, $T_{\mathrm{i}}^{\prime }=10^{9}$ K, $\protect\kappa =1~%
\mathrm{cm^{2}\ g^{-1}}$, $\protect\kappa _{\mathrm{X}}=0.05~\mathrm{cm^{2}\
g^{-1}}$, $\protect\epsilon _{\mathrm{e,rs}}=0.9$, electron and magnetic
equipartition factors of forward shock ($\protect\epsilon _{\mathrm{e}}=0.1$%
, $\protect\epsilon _{\mathrm{B}}=0.01$). Other plots depict the light
curves with one or two parameters altered. Panel (b) $L_{\mathrm{sd}}=10^{49}~%
\mathrm{erg\ s^{-1}}$, $t_{\mathrm{sd}}=10^{3}$ s. Panel (c) $M_{\mathrm{ej}%
}=10^{-2}~M_{\odot }$. Panel (d) $\protect\gamma _{4}=10^{5}$.}
\label{fig: comparison_paras}
\end{figure}

In the following, we study the emission features of the three
regions and their dependence on some key parameters, based on comparison of
the plots arranged in Figure~\ref{fig: comparison_paras}.

We take the set of parameters used in Figure~\ref{fig:
comparison_paras}(a) as a control group and change one or two to see the
difference. As X-ray light curves in Figure~\ref{fig: comparison_paras}(a)
show, the thermal emission peaks before the reverse-shock emission, while the forward
shock emission dominates after $\sim 10^{6}$ s. As for optical band, the
reverse-shock emission dominates over the thermal emission, although they
peak simultaneously. Figure~\ref{fig: comparison_paras}(b) implies that
faster energy injection leads to earlier, higher peaks of the reverse-shock
emission and later, weaker peaks of the forward shock emission. As seen in
Figure~\ref{fig: comparison_paras}(c), the more massive the merger ejecta,
the more important the thermal emission at optical band compared with the
reverse-shock emission. According to Figure~\ref{fig: comparison_paras}(d), $%
\gamma _{4}$ only affects the components of reverse shock. In the case of
higher $\gamma _{4}$, the reverse-shock emission reaches a lower peak value and
then has a flatter decay.

Based on the multiwavelength behaviors of the thermal emission, we find
that the X-ray peak always occurs much earlier than the optical peak.
However, the X-ray peak of the reverse-shock emission is followed closely by the
optical peak. In some cases, the thermal and reverse-shock emission at
optical band peaks nearly simultaneously. Thus, this model can reproduce
late bumps at different wavebands which occur almost at the same time.

\section{Application}

\label{sec: application}

As can be seen in the fitting results of
\citet{Gao2015ApJ...807..163G,
Gao2017ApJ...837...50G}, the afterglows of both GRB 080503 and GRB 050724
show the concurrence between optical/IR and X-ray bumps. For GRB 080503,
however, since there are no confirmed redshift and fewer observational data,
we choose GRB 050724 as the object for model fitting. Burst Alert Telescope
on board the \textit{Swift} satellite, at 12:34:09 UT on July 24th in 2005,
triggered and detected GRB 050724 dominated by an initial hard peak lasting
for 0.256 s \citep{Covino2005, Krimm2005}. Despite the duration $T_{90}>2$ s
given by \textit{Swift}, there are three pieces of evidence supporting that
this burst can be classified as an sGRB \citep{Barthelmy2005}. Firstly, the
duration derived from the 50keV-350keV data is consistent with the typical
sGRB property. Secondly, the elliptical galaxy harboring GRB 050724 (at a
redshift of 0.257; \citealt{Berger2005}) is a proper location for binary
compact star mergers, which is the most promising progenitors of sGRBs.
Lastly, the isotropic energy of the prompt emission is $E_{\gamma,\mathrm{%
iso}}\sim 10^{50}$ $\mathrm{erg}$, which is $2\sim 3$ orders of magnitude
lower than that of typical long GRBs.

\begin{figure}[tbp]
\centering\resizebox{1\hsize}{!}{
\includegraphics{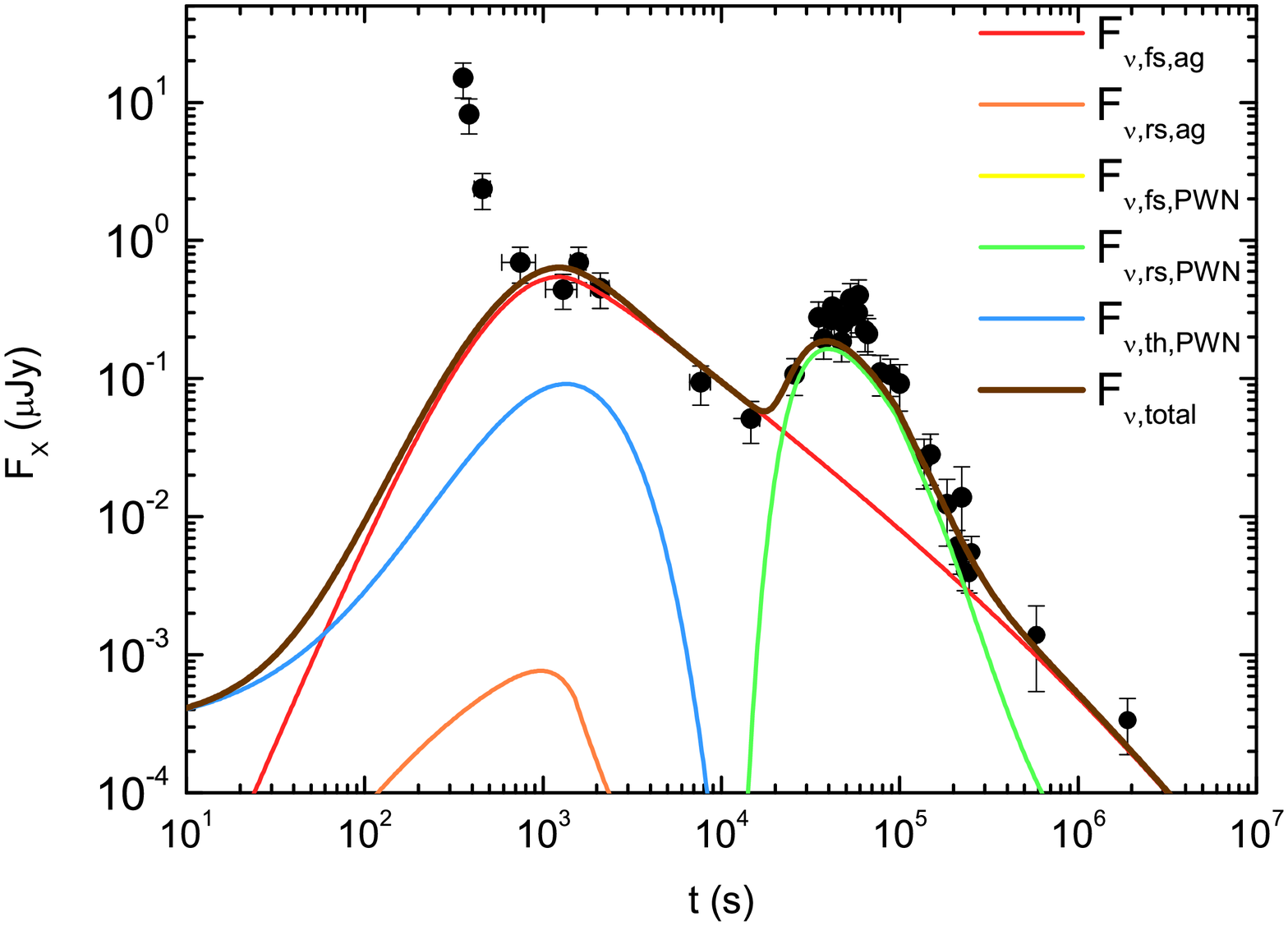}}
\resizebox{1\hsize}{!}{
\includegraphics{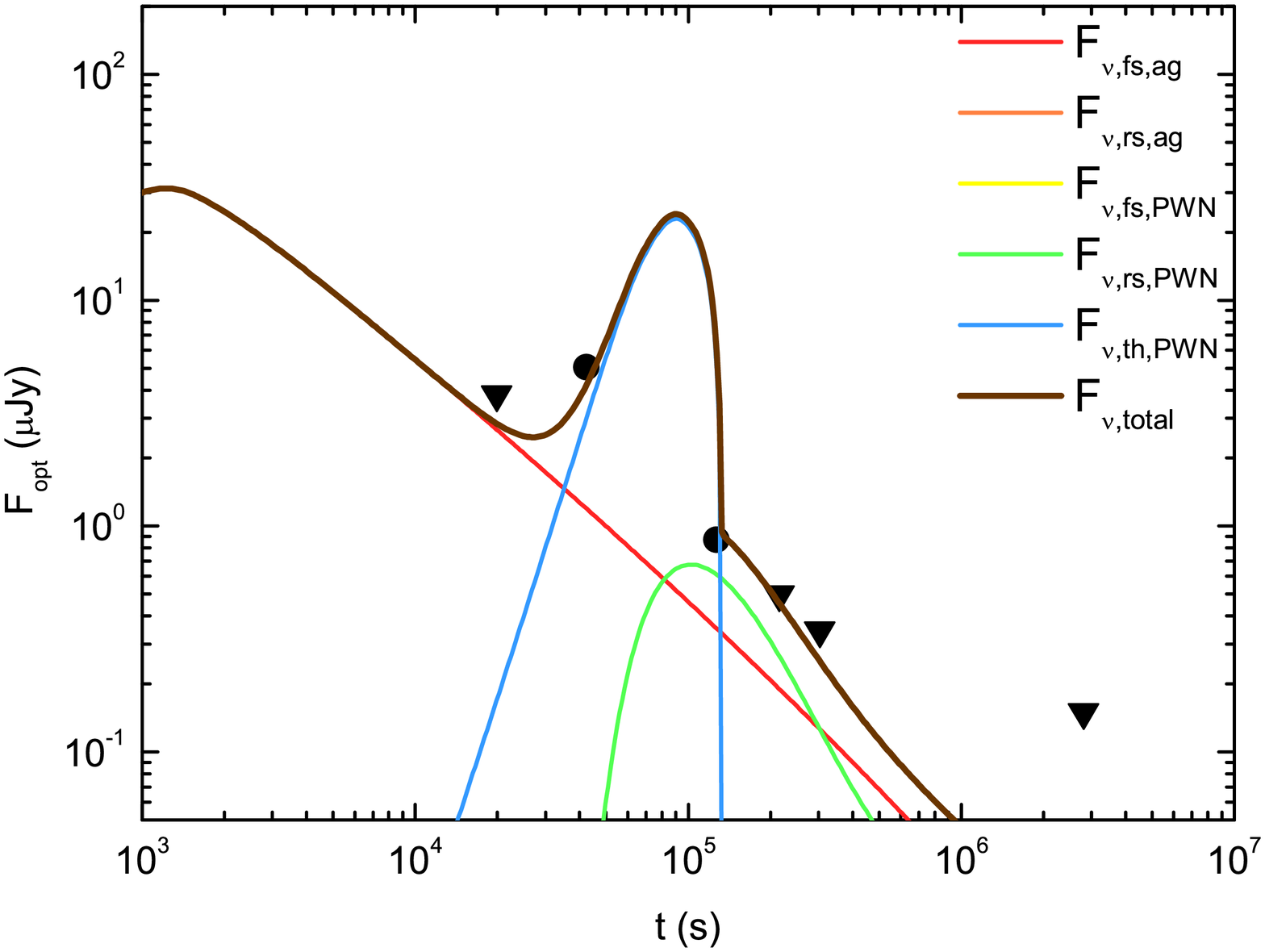}}
\resizebox{1\hsize}{!}{
\includegraphics{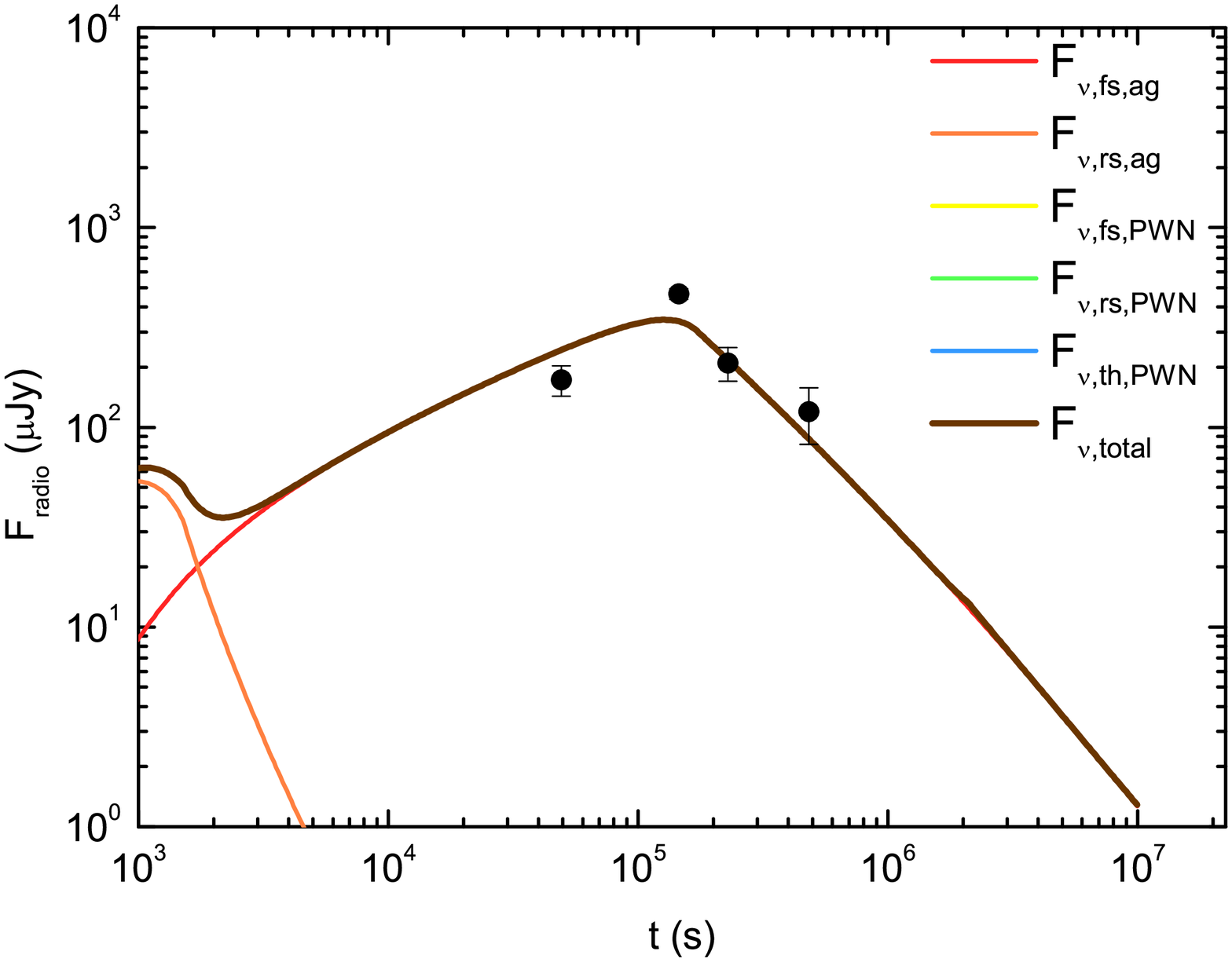}}
\caption{Fitting results to the broadband data of GRB 050724. The correction
for Galactic or intrinsic extinction has not been applied to the optical
data. Displayed are the contributions from the jet forward shock (red), jet
reverse shock (orange) and merger ejecta system radiation comprised of
the forward-shock emission (yellow), thermal emission (green) and reverse-shock emission (blue).
The total flux is denoted as brown lines. It is shown that the emission from the reverse
shock of the jet and the forward shock in the nebula is not notable.}
\label{fig: GRB050724}
\end{figure}

The observational data of GRB 050724 in X-ray, $R$ and radio (8.46 GHz)
bands are collected from the literature
\citep{Berger2005, Burenin2005GCN..3671....1B, Grupe2006, Panaitescu2006MNRAS.367L..42P,
Malesani2007A&A...473...77M}. The most prominent feature of the
light curves of GRB 050724 is the optical bump and the bright X-ray flare
occurred at approximately 1 day after the GRB trigger, as can be seen in
Figure~\ref{fig: GRB050724}. Several models have been proposed to interpret
this feature
\citep{Panaitescu2007MNRAS.379..331P, Gao2015ApJ...807..163G,
Gao2017ApJ...837...50G}.

Here we propose that the optical bumps and X-ray flares at several days
after the GRB trigger are related to PWN emission powered by a magnetar. The
fitting results for GRB 050724 are shown in Figure~\ref{fig: GRB050724} with
adopted parameters presented in Table~\ref{tab: parameters}.

Our fitting results are summarized in what follows. Our model suggests that
the thermal radiation of the merger ejecta is responsible for the optical
bump at $\sim 1$ day, in agreement with the kilonova model. The
reverse-shock emission from behind the merger ejecta proves to be capable of
producing a late bright X-ray flare. Here we suggest that the optical and
X-ray re-brightenings both occurring at days after GRB trigger may not share
the same origin, which is supported by the observations that optical bumps
are not always accompanied by X-ray flares. The normal decay in the X-ray
band after the flare, attributed here to the non jet-break before $\sim
10^{6}$ s since burst, requires relatively large opening angle of the jet,
which is consistent with the low degree of collimation suggested by %
\citet{Grupe2006}.

\begin{table*}[tbp]
\caption{Adopted parameters for fitting the broadband observations of GRB
050724.}
\label{tab: parameters}\centering%
\begin{tabular}{cccccccc}
\hline\hline
\multicolumn{8}{c}{Jet parameters} \\
$E_{\mathrm{iso,\mathrm{j}}}$ ($\mathrm{erg}$) & $\theta _{\mathrm{j}}$ (rad)
& $\gamma _{4,\mathrm{j}}$ & $\Delta_{\mathrm{4,i}}$ (cm) &  &  &  &  \\
\hline
$5\times 10^{51}$ & 0.4 & 80 & $10^{12}$ &  &  &  &  \\ \hline\hline
\multicolumn{8}{c}{PWN parameters} \\
$L_{\mathrm{sd}}$ ($\mathrm{erg\ s^{-1}}$) & $t_{\mathrm{sd}}$ (s) & $M_{%
\mathrm{ej}}$ ($M_{\odot }$) & $v_{i}/c$ & $\gamma _{4}$ & $T_{\mathrm{i}%
}^{\prime }$ (K) & $\kappa$ ($\mathrm{cm^{2}\ g^{-1}}$) & $\kappa _{\mathrm{X%
}}$ ($\mathrm{cm^{2}\ g^{-1}}$) \\ \hline
$3\times 10^{46}$ & $2\times 10^{5}$ & $10^{-2}$ & 0.2 & $10^{5}$ & $%
10^{9} $ & 1 & 0.1 \\ \hline\hline
\multicolumn{8}{c}{Other parameters} \\
$n_{1}$ ($\mathrm{cm^{-3}}$) & $p$ & $\epsilon _{\mathrm{e}}$ & $\epsilon _{%
\mathrm{B}}$ & $\epsilon _{\mathrm{e,rs}}$ &  &  &  \\ \hline
0.0005 & 2.3 & 0.1 & 0.01 & 0.9 &  &  &  \\ \hline
\end{tabular}%
\end{table*}

In the following, we make a detailed analysis of the adopted parameters. The
afterglow X-ray light curve with the decay slope $\alpha _{\mathrm{X}%
}=0.98_{-0.09}^{+0.11}$ \citep{Grupe2006}, together with radio data, places
a tight constraint on the electron spectral index $p$.\footnote{%
In principle the electron spectral indices of the forward- and
reverse-shocked electrons can be different. In reality we found that the
difference of adopting different values for these two parameters is small
when fitting the observational data.} However, there are only loose
constraints on other parameters associated with the afterglow component,
i.e., the isotropic kinetic energy ($E_{\mathrm{iso,\mathrm{j}}}$), half
opening angle ($\theta _{\mathrm{j}}$), Lorentz factor of region 4 ($\gamma
_{4,\mathrm{j}}$), initial width of region 4 ($\Delta_{\mathrm{4,i}}$) and
electron and magnetic equipartition factors ($\epsilon _{\mathrm{e}}$, $%
\epsilon _{\mathrm{B}}$). Note that $\epsilon _{\mathrm{e}}$ and $\epsilon _{%
\mathrm{B}}$ are applied here in shock emission of all other cases except
for the reverse-shocked magnetar wind which uses $\epsilon _{\mathrm{e,rs}}$
and $\epsilon _{\mathrm{B,rs}}=1-\epsilon _{\mathrm{e,rs}}$.

In the modeling of the magnetar-powered $r$-process ejecta, there is some
uncertainty in the values of the ejecta opacities in the optical band $%
\kappa $ and the X-ray band $\kappa _{\mathrm{X}}$ \citep{Wang+Dai2015}.
During the time ($\sim 1$ day) when the optical bump of GRB 050724 is
prominent, the ejecta temperature is $\gtrsim 10^{5}$ K. At such
temperatures, the bound-bound opacity is negligible %
\citep{Kasen2013ApJ...774...25K}. According to the analysis of \cite%
{Wang+Dai2015}, the optical opacity of the ejecta is dominated by the
bound-free opacity, which we estimate to be $\kappa =1~\mathrm{cm^{2}\ g^{-1}%
}$. For opacity in X-ray band, we set $\kappa _{\mathrm{X}}=0.1~\mathrm{%
cm^{2}\ g^{-1}}$, which is close to the electron scattering opacity $\kappa
_{\mathrm{es}}\approx 0.01~\mathrm{cm^{2}\ g^{-1}}$ \citep{Wang+Dai2015}. We
choose such a value that is slightly larger than $\kappa _{\mathrm{es}}$ to
account for the possible contribution from the bound-free transitions. We
note that the resulting light curves are not sensitive to the initial
temperature of the ejecta $T_{\mathrm{i}}^{\prime }$.

From the values of $L_{\mathrm{sd}}$ and $t_{\mathrm{sd}}$, we can derive
the properties of the central magnetar, i.e., surface magnetic field $%
B=2.7\times 10^{14}$ G and initial spin period $P_{0}=2.2$ ms, being
consistent with the proposal of millisecond magnetars as likely products
from the coalescence of binary NSs. Furthermore, $t_{\mathrm{sd}}=2\times
10^{5}$ s suggests a stable magnetar. The merger ejecta with mass $M_{%
\mathrm{ej}}=10^{-2}\ M_{\odot }$ and initial speed of $v_{i}=0.2c$ meet the
numerical simulations of NS-NS mergers, i.e., $M_{\mathrm{ej}}\sim
10^{-4}-10^{-2}\ M_{\odot }$ and $v_{i}\sim 0.1-0.3c$ %
\citep{Hotokezaka2013PhRvD..87b4001H}. The above results suggest that
magnetar formation following NS-NS mergers is consistent with at least a
subset of sGRBs.

\section{Conclusions}

\label{sec: summary}

Several sGRBs are found to have prominent features of near-infrared/optical
re-brightenings and X-ray flares at days after GRB triggers. There features
are not expected in the standard afterglow model and are therefore a direct
reflection of the continuous activity of the central engine.

In this paper we propose that these deviations from the standard afterglow
model can be interpreted by the formation of stable magnetars following the
mergers of double NSs. The rapidly-spinning magnetar blows up a nebula, as
schematized in Figure \ref{fig:sketch}. In describing the PWN model modified
from \citet{Wang+Dai2015}, we stress the dynamic effect of significant
radiation energy that is emitted by reverse-shocked $e^{+}e^{-}$ pairs and
strongly absorbed by the ejecta in the early time. The radiation energy loss
or transfer weaken substantially the acceleration by the magnetar wind.

Based on the fitting results of the multi-band data of GRB 050724, we make
main conclusions as follows. Firstly, a good fit to sGRB observations with
the PWN model supports the formation of long-lived magnetars after NS-NS
mergers. Secondly, late optical bumps are attributed to thermal radiation
from merger ejecta which are heated by both the $r$-process material and
reverse-shock radiation at early times. Thirdly, different from the origin
of optical bumps, X-ray flares occurring at $\sim 1$ day are likely to
result from the synchrotron emission of reverse-shocked $e^{+}e^{-}$ wind.
Finally, due to the notable properties of nebula emission, we suggest that
PWNe could be an efficient probe for the NS-NS mergers and newborn
millisecond magnetars.

The breakthrough detection of GW 170817 accompanied by rich
electromagnetic counterparts including sGRB 170817A boosts the researches relevant
to sGRBs and marks the beginning of a new era of multi-messenger astronomy. Future
combined detections of NS-NS coalescences with total mass consistent with
stable NSs and electromagnetic counterparts with PWN characteristics would
further provide strong evidence for the PWN model presented in this paper.

\acknowledgments

We thank Fang-Kun Peng for his helpful discussion about the \textit{Swift}
data. This work was supported by the National Basic Research Program
(\textquotedblleft 973" program) of China (grant No. 2014CB845800), the
National Natural Science Foundation of China (grant No. 11573014) and the National Key R\&D Program of China (grant no. 2017YFA0402600). L.J.W.
was also supported by the National Program on Key Research and Development
Project of China (Grant No. 2016YFA0400801).

\appendix

\section{A. Calculation of quantities in GRB jets}

\label{app: calculation of jets}

According to the mass conservation on both sides of the reverse-shock surface,
the width of regions 3 and 4 increases by
\begin{equation}
\mathrm{d} \Delta_{3,\mathrm{j}} = \dfrac{(\beta_{4,\mathrm{j}}-\beta_%
\mathrm{j})/\beta_{\mathrm{fs,\mathrm{j}}}}{\dfrac{\gamma_\mathrm{j} n_{3,%
\mathrm{j}}^{\prime}}{\gamma_{4,\mathrm{j}} n_{4,\mathrm{j}}^{\prime}}-1}%
\mathrm{d} r_\mathrm{j}  \label{eq: region 3 width}
\end{equation}
\begin{equation}
\mathrm{d} \Delta_{4,\mathrm{j}}=-\dfrac{\gamma_{\mathrm{j}}n_{3,\mathrm{j}%
}^{\prime}}{\gamma_{4,\mathrm{j}} n_{4,\mathrm{j}}^{\prime}} \mathrm{d}
\Delta_{3,\mathrm{j}} ,  \label{eq: region 4 width}
\end{equation}
where $\beta_{\mathrm{fs,\mathrm{j}}}= \beta_\mathrm{j}/\left(1-1/(4\gamma_%
\mathrm{j}+3)/ \gamma_\mathrm{j}\right) $ is the forward shock velocity in
units of $c$ and the number density ratio of region 3 and 4, $n_{3,%
\mathrm{j}}^{\prime}/n_{4,\mathrm{j}}^{\prime}=4\gamma_{34,\mathrm{j}}+3$
satisfies the shock jump conditions %
\citep{Blandford+McKee1976PhFl...19.1130B}.

The other quantities about GRB jets evolve as
\begin{equation}
\dfrac{\mathrm{d} r_\mathrm{j}}{\mathrm{d} t}= \dfrac{\beta_{\mathrm{fs,%
\mathrm{j}}} c}{1-\beta_{\mathrm{fs,\mathrm{j}}}},  \label{eq: dR_ag}
\end{equation}
\begin{equation}
\dfrac{\mathrm{d} \theta_{\mathrm{j}}}{\mathrm{d} r_\mathrm{j}}= c_{\mathrm{s%
}} \mathcal{D}/r_\mathrm{j},  \label{eq: dtheta_ag}
\end{equation}
\begin{equation}
\dfrac{\mathrm{d} M_{\mathrm{sw},\mathrm{j}}}{\mathrm{d} r_\mathrm{j}}= 4\pi
(1-\cos \theta_\mathrm{j})r_\mathrm{j}^2 n m_{\mathrm{p}},
\label{eq: dMsw_ag}
\end{equation}
\begin{equation}
\dfrac{\mathrm{d} M_{3,\mathrm{j}}}{\mathrm{d} r_\mathrm{j}}=4\pi (1-\cos
\theta_\mathrm{j}) r_\mathrm{j}^2 n_3^{\prime} m_{\mathrm{p}}\gamma_\mathrm{j%
}\dfrac{\mathrm{d} \Delta_{3,\mathrm{j}}}{\mathrm{d} r_\mathrm{j}},
\label{eq: dM3_ag}
\end{equation}
\begin{equation}
n^{\prime}_{4,\mathrm{j}}=\dfrac{M_{\mathrm{ej,j}}}{4\pi(1-\cos \theta_%
\mathrm{j})r_{\mathrm{ej}}^2 \gamma_4 \Delta_4 m_{\mathrm{p}}},
\label{eq: dn4_ag}
\end{equation}
where $r_{\mathrm{j}}$ is the radius, $\theta _{\mathrm{j}}$ is the half
opening angle of jets and $n_{4,\mathrm{j}}^{\prime }$ is the comoving-frame
number density of particles in region 4. Note that the sound speed $%
c_{\mathrm{s}}$ is calculated according to Equation (10) of \citet{Huang2000}%
, which is applicable for both relativistic and non-relativistic scenarios.

\section{B. The internal energy gained by the reverse shock in the PWN}

\label{app: dUrs}

\begin{figure*}[tbp]
\resizebox{0.5\hsize}{!}{
\includegraphics{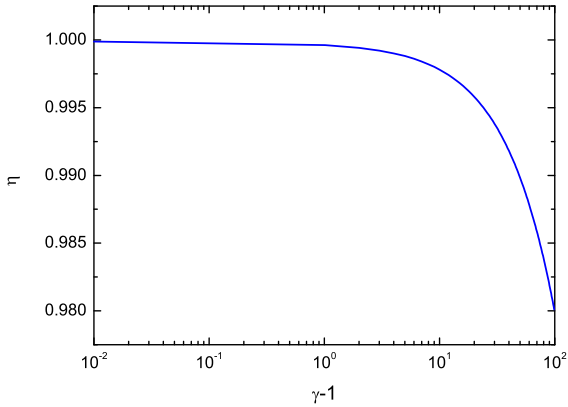}}
\caption{$\protect\eta $ as a function of $\protect\gamma -1$.}
\label{fig: eta}
\end{figure*}

When typical parameters for PWNe are adopted, i.e., $\gamma _{4}=10^{4}$ and
$\beta \sim 0.1$, we have the width of region 3 $\mathrm{d}\Delta _{3}\sim
\mathrm{d}r$, leading to the conclusion that the reverse-shock surface stands
near the central magnetar. So we can obtain $\mathrm{d}M_{3}/\mathrm{d}t=L_{%
\mathrm{sd}}/\left( \gamma _{4}c^{2}\right) $ and then derive
\begin{equation}
\dfrac{\mathrm{d}U_{\mathrm{3,rs}}^{\prime }}{\mathrm{d}t^{\prime }}=\dfrac{%
\gamma _{34}-1}{\gamma _{4}}\mathcal{D}L_{\mathrm{sd}}^{\prime }.
\label{eq: dUrs}
\end{equation}%
By defining $\eta \equiv \dfrac{\mathrm{d}U_{\mathrm{3,rs}}^{\prime }/%
\mathrm{d}t^{\prime }}{L_{\mathrm{sd}}^{\prime }}=\dfrac{\gamma _{34}-1}{%
\gamma _{4}}\mathcal{D}$, the relation between $\eta $ and $\gamma -1$ is
shown in Figure~\ref{fig: eta}, which suggests $\dfrac{\mathrm{d}U_{\mathrm{%
3,rs}}^{\prime }}{\mathrm{d}t^{\prime }}\sim L_{\mathrm{sd}}^{\prime }$.

\end{document}